\documentclass[preprint,preprintnumbers,amsmath,amssymb,nofootinbib,aps,usenatbib]{revtex4}
\usepackage[T1]{fontenc}

\usepackage{float,amsmath, amssymb,epsfig,graphicx,mathrsfs}
\usepackage{amsfonts,bbm,color,bm}
\usepackage{amsmath}
\usepackage[colorlinks]{hyperref}
\usepackage[figure,table]{hypcap}
\usepackage[textsize=tiny]{todonotes}
\usepackage{subfigure}

\def\aap{Astron. Astrophys.}
\def\mnras{Mon. Not. Roy. Astron. Soc.}
\def\apjl{Astrophys. J. Lett.}
\def\jcap{JCAP}
\def\prd{Phys. Rev. D}
\def\apj{Astrophys. J.}

\def\nar{New Astronomy Reviews}
\def\pasp{Publications of the Astronomical Society of the Pacific}
\def\epjc{Eur. Phys. J. C}
\newcommand{\Mpc}{\mathrm{~km~s^{-1}~Mpc^{-1}}}
\begin{document}

\title{Revisiting Friedmann-like cosmology with torsion: newest constraints from high-redshift observations}

\author{Tonghua Liu$^{1,2}$}
\author{Ziqiang Liu$^{1}$}
\author{Jiamin Wang$^{1}$}
\author{Shengnan Gong$^{1}$}
\author{Man Li$^{1}$}
\author{Shuo Cao$^{2,3}${\footnote{Corresponding author: caoshuo@bnu.edu.cn}}}

\affiliation{
1. School of Physics and Optoelectronic, Yangtze University, Jingzhou 434023, China;\\
2. Institute for Frontiers in Astronomy and Astrophysics, Beijing Normal University, Beijing 102206, China;\\
3. Department of Astronomy, Beijing Normal University, Beijing 100875, China.}

\baselineskip=0.68 cm
\vspace*{0.25cm}
\begin{abstract}
As one of the possible extensions of Einstein's General Theory of Relativity, it has been recently suggested that the presence of space-time torsion could solve problems of the very early and the late-time universe undergoing accelerating phases. In this paper, we use the latest observations of high-redshift data, coming from multiple measurements of quasars and baryon acoustic oscillations, to phenomenologically constrain such cosmological model in the framework of Einstein-Cartan (EC) endowed with space-time torsion. Such newly compiled quasar datasets in the cosmological analysis is crucial to this aim, since it will extend the Hubble diagram to high-redshift range in which predictions from different cosmologies can be distinguished. Our results show that out of all the candidate models, the torsion plus cosmological constant model is strongly favoured by the current high-redshift data, where torsion would be expected to yield the late-time cosmic acceleration. Specially, in the framework of Friedmann-like cosmology with torsion, the determined Hubble constant is in very good agreement with that derived from the Planck 2018 CMB results. On the other hand, our results are compatible with zero spatial curvature and there is no significant deviation from flat spatial hypersurfaces. Finally, we check the robustness of high-redshift observations by placing constraints on the torsion parameter $\alpha$, which is strongly consistent with other recent works focusing on torsion effect on the primordial helium-4 abundance.

\end{abstract}

\vspace*{0.25cm}
\maketitle

\section{Introduction}\label{sI}

The cosmological constant plus cold dark matter model (called the $\Lambda$CDM model) is considered the standard cosmological model because it has withstood most popular observational evidences, such as the observations of type Ia supernovae (SNe Ia) \citep{2007ApJ...659...98R,Alam17}, Cosmic Microwave Background Radiation (CMB) \citep{2016A&A...594A..13P}, and strong gravitational lensing \citep{cao2012SL,cao2014cosmic,Cao:2015qja,2022ChPhL..39k9801L,Liuapj1}. However, the standard cosmological model has been suffering several problems, such as the \textit{coincidence} problem and the well-known \textit{fine-tuning} problem \citep{1989RvMP...61....1W,2000astro.ph..5265W,2001PhRvL..87n1302D,2010ApJ...711..439C,2011MNRAS.416.1099C,2018RAA....18...66Q}.
With the rapid development of observational techniques, astronomical data are becoming more accurate,
more arguments are emerging for $\Lambda$CDM model. The relatively significant two issues are the Hubble tension \citep{2017NatAs...1E.169F,2021APh...13102605D} and the cosmic curvature problems \citep{2020NatAs...4..196D}. More specifically, there is $4.4\sigma$ tension between the Hubble constant ($H_0$) inferred within $\Lambda$CDM from cosmic microwave background (CMB) anisotropy (temperature and polarization) data \citep{2020A&A...641A...6P} and the local measured $H_0$ from  using calibrated distance ladder method by \textit{Supernova $H_0$ for the Equation of State} collaboration (SH0ES) \citep{2019ApJ...876...85R}. The discrepancy suggests either an unknown systematic error in astrophysical observations or new physics significantly different from $\Lambda$CDM. Some recent works have suggested that this discrepancy may be due to inconsistencies in the spatial structure of the early and late Universe \citep{2021APh...13102605D,2020NatAs...4..196D,2021PhRvD.103d1301H}. The Planck lensing data plus low redshift baryon acoustic oscillations observations have shown the Universe is quite flat  with curvature parameter precisely constrained within $\Omega_k=0.001\pm0.002$ \citep{2020A&A...641A...6P}.  However, a closed Universe $\Omega_k=-0.044^{+0.018}_{-0.015}$ was supported by using the \textit{Planck} temperature and polarization power spectra data alone due to as a consequence of higher lensing amplitude \citep{2020NatAs...4..196D}. It should be emphasized that both methods invoke a particular cosmological model --- the non-flat $\Lambda$CDM. A review for challenges to the standard $\Lambda$CDM model please see \citep{2022NewAR..9501659P}.

Since there is no evidence of considerable systematic errors in Planck observations and local measurements \citep{2019ApJ...876...85R,DiValentino:2018zjj,Follin:2017ljs}, there has been increasing attention to other cosmological models beyond $\Lambda$CDM, such as Early Dark Energy models \citep{Tanvi} and interacting dark energy models that consider dark energy interacting with dark matter \citep{Valiviita,ZXG,Li17}. Some recent works showed that the generalized Chaplygin gas model could greatly reduce the tension between the CMB and local determinations of the Hubble constant \citep{Yang}. Moreover, a Phenomenologically Emergent Dark Energy (PEDE) presented by Li \textit{et al.} has obtained a great attention \citep{Liam19}, which demonstrated its potential in addressing the Hubble constant problem. Furthermore, modifications to General Relativity (GR) theory \citep{Hashim1,Hashim2,Briffa,Ren} or the well-known Dvali-Gabadadze-Porrati model physically motivated by possible multidimensionality in the brane theory provides another way to deal with the cosmological constant problem and alleviate the Hubble tension \citep{Cao:2017ivt,Cao2017JCAP,Xu10,Xu14,Giannantonio08,Liu11,Liu12}. Note that all of these models could not only explain the late-time cosmic acceleration from different mechanisms, but also describe the large-scale structure distribution of the Universe (see \citep{Koyama16} for recent reviews). However, it's still doubtful that dark energy or dark matter exists at all. In recent years, an extension of general relativity theory, the Einstein-Cartan (EC) theory with the presence of space-time torsion has gained a lot of attention, where torsion acts a new dark source of torsionless Riemannian gravity and drives the expansion of Universe (see \cite{BH} for a recent review). Some works have even proposed that dark matter could be torsion in disguise. Although no experimental or observational evidence to support the distinctive predictions of the EC theory \cite{1,2,3,4,5,6}, the main reason is that the theory only deviates from classical general relativity at extremely high energy densities. These densities can only be reached deep inside dense objects, such as neutron stars and black holes, or during the early stages of the expansion of Universe. Nevertheless, EC theory can still provide some answers to unsolved problems in modern theoretical physics and astrophysics, such as
the singularity question, inflationary models \cite{7,8}, the recently discovered universal acceleration \cite{9,10}, and what is required of any bridge between general relativity and quantum mechanics. Please see reference \cite{ECcs} for recent review of EC cosmological research.

In this work, we will explore the validity of a cosmological model in an EC framework endowed with torsion, focusing on the angular size measurements of ultra-compact structure in radio quasars as standard rulers and nonlinear relation between ultraviolet (UV) and X-ray luminosities of quasar acting standard candles at redshift covering $0.009<z<7.5$. Additionally, we also adopt 11 data points 
of Baryon Acoustic Oscillation (BAO) from BOSS DR12 at $z=0.38, 0.51, 0.61$, 6dFGs and SDSS MGS at $z=0.122$, DES Y1 results at $z=0.81$, and eBOSS DR14 at $z=1.52, 2.34$. Specially, the Hubble constant, spatial curvature and strength of torsion field parameter $\lambda$ will be visited with such combination of two types of newly compiled quasar samples. The outline of this paper is given as follow: In section 2 we introduce the Einstein-Cartan equation of gravitation and Friedmann equations in a Friedmann-Lematre-Robertson-Walker (FRLW) Universe. The observational two types of newly compiled high redshift quasar datasets are given in section 3. The methodology and results are shown in section 4. Finally, we summarise our conclusions in section 5. We adopt $c = k_B = 1$ units throughout this work.

\section{Friedmann-like cosmology with torsion}

Extensions of General Relativity beyond the boundaries of Riemannian geometry, by allowing asymmetric affine connections, have a long history in the literature. These studies introduce the possibility of space-time torsion and associate new degrees of freedom into the gravitational field (e.g.~see~\cite{BH} for a recent review). Therefore, it comes to no surprise that there are many applications of these theories for cosmology, in an effort to illuminate the role and the potential implications of torsion (as well as those of spin) for the evolution ofthe Universe.

The EC theory was first proposed by Cartan in 1922, based on a simple physical intuition, that torsion effect is regarded as a macroscopic manifestation of the intrinsic angular momentum (spin) of matter \cite{11}. Kibble and Sciama then reintroduced the spin of matter independently into GR also known as ECKS theory \cite{12}. The ECKS theory postulates an asymmetric affine connection for the space-time, in contrast to the symmetric Christoffel symbols of Riemannian spaces. In technical terms, torsion is described by the antisymmetric part of the non-Riemannian affine connection. Here we give a brief introduction to the torsion theory, and further give the Friedman equation and continuity equation.

\subsection{space-time with torsion}

With the presence of torsion terms into Einstein gravitational field, it is known as the Einstein-Cartan equation of gravitation \cite{Kranas2019}. The Einstein-Cartan field equation of gravitation describes the relationship between matter and the geometry of space-time, which combines Ricci tensor, Ricci scalar and energy momentum tensor
\begin{equation}
R_{\mu\nu} - \frac{1}{2}R g_{\mu\nu}=\kappa T_{\mu\nu}-\Lambda g_{\mu\nu} \,,\label{Rmunu}
\end{equation}
where $\kappa = {8\pi G}$ and $\Lambda$ is the cosmological constant. Considering the space-time with non-zero torsion, the affine connection contains an antisymmetric part, $\Gamma^\alpha_{~\mu\nu}=\Tilde{\Gamma}^\alpha_{~\mu\nu}+K^\alpha_{~\mu\nu}$, where $\Tilde{\Gamma}^\alpha_{~\mu\nu}$ is the symmetric Christoffel symbol, and $K^\alpha_{~~\mu\nu}$ is the contortion tensor given by
\begin{equation}
        K^\alpha_{~\mu\nu} = S^\alpha_{~\mu\nu} + S_{\mu\nu}^{~~\alpha} + S_{\nu\mu}^{~~\alpha}. \label{K}
\end{equation}
The antisymmetry of the torsion tensor $S^\alpha_{~~\mu\nu}$ guarantees that it has only one non-trivial contraction, leading to the torsion vector $S_{\mu}=S^\nu_{~\mu\nu} = - S^\nu_{~\nu\mu}$. From the physical point of view, torsion can be regarded as a possible link between the intrinsic angular momentum of matter (known as spin), and the geometry of the main space-time \cite{Pasmat2017}. 

In the space-time with torsion, the energy momentum tensor is also antisymmetric and typically coupled to the spin of the matter via the Cartan field equations. The spin tensor is given by
\begin{equation}
        S_{\alpha\mu\nu} = -\frac{1}{4}\kappa (2s_{\mu\nu\alpha}+g_{\nu\alpha}s_\mu - g_{\alpha\mu}s_\nu)\,, \label{Sss}
\end{equation}
where $s_\alpha=s^\mu_{~~\alpha\mu}$ is the spin vector of the matter. In a homogeneous and isotropic Friedmann background, the torsion tensor and the associated vector have the form \cite{Kranas2019}
\begin{equation}
     S_{\alpha\mu\nu} = \phi (h_{\alpha\mu} u_\nu - h_{\alpha\nu}u_\mu),   \hspace{1cm}  S_\alpha = -3\phi u_\alpha\,, \label{SS}
\end{equation}
where $\phi=\phi(t)$ is the scalar function that depends only on time due to homogeneity of space,  $h_{\mu\nu}=g_{\mu\nu}+u_{\mu} u_{\nu}$ is a  symmetric projection tensor which is orthogonal to the 4-vector velocity $u_\mu$.

\subsection{Friedmann equation in torsion cosmology with cosmological constant }

\label{sec:torsion}
The foundation of modern cosmology is based on the basic principles of cosmology, i.e., the Universe is homogeneous and isotropic at large scales \cite{Cao2019SR,Qi2019PRD}. The Friedmann-Lematre-Robertson-Walker (FLRW) metric which is appropriate to describe 
a homogeneous and isotropic universe, read as 
\begin{equation}
ds^2=dt^2-\frac{a(t)^2}{1-kr^2}dr^2-a(t)^2r^2d\Omega^2,
\end{equation}
where $a(t)$ is the scale factor, and $k$ is dimensionless curvature taking one of three values $\{-1, 0, 1\}$ corresponding to a close, flat and open universe.

In the presence of torsion, the analogues of the Friedmann and the Raychaudhuri equations, in a space-time with nonzero spatial curvature and non-vanishing cosmological constant (i.e.~when $k,\Lambda\neq0$) take the form \cite{Kranas2019} 
\begin{equation}
\left({\frac{\dot{a}}{a}}\right)^{2}= {\frac{1}{3}}\,\kappa\rho- \frac{k}{a^{2}}+ {\frac{1}{3}}\,\Lambda- 4\phi^{2}- 4\left({\frac{\dot{a}}{a}}\right)\phi  ,\label{Fried1}
\end{equation}
and
\begin{equation}
{\frac{\ddot{a}}{a}}= -{\frac{1}{6}}\,\kappa\left(\rho+3p\right)+ {\frac{1}{3}}\,\Lambda- 2\dot{\phi}- 2\left({\frac{\dot{a}}{a}}\right)\phi\,,  \label{Ray1}
\end{equation}
where $\rho$ and $p$ are the energy density and pressure of matter.
With the definition of the equation of state $w=p/\rho$, the continuity equation is given by the following form \cite{Kranas2019}
\begin{equation}
\dot{\rho}+3(1+w)H\rho + 2(1+3w)\phi \rho = 4\phi{\Lambda}{\kappa^{-1}}\,,\label{rhodot}
\end{equation}
where $H=\dot{a}/a$ is defined as the Hubble parameter. Note that the above equation recovers the standard  continuity equation 
when $\phi=0$.

\subsection{Case I: $\phi(t)=-H(t)/2$  with vanishing cosmological
constant}
Considering a constant and negative $\phi$ plus vanishing cosmological
constant case, torsion tends to accelerate the cosmic expansion. If assuming a flat and empty space ($k=\rho=\Lambda=0$), Eq. (\ref{Fried1}) will become $\phi(t)=-H(t)/2$, which means the torsion function $\phi(t)$ depends on Hubble parameter $H$. Thus, we adopt a simple parameterized form $\phi(t)=-\alpha H(t)$ with $\Lambda=0$ and $\omega=0$ (assuming that the matter content is dark matter or baryonic matter). The analytic solution of Eq.~(\ref{rhodot}) is
\begin{equation}
    \rho_m(a)=\rho_{0m}\bigg(\frac{a}{a_0}\bigg)^{-3+2\alpha},
\end{equation}
and the Friedmann equation becomes
\begin{equation}
    H^2=\frac{\kappa}{3}\rho_m - \frac{k}{a^2}+4\alpha  H^2 - 4 \alpha^2 H^2.\ \label{H2case2}
\end{equation}
\textbf{This Equation clearly shows that torsion contributes to the total effective energy-density of the system. Similar to the FLRW Universe model, the density parameter $\Omega$ can also be introduced for a homogeneous and isotropic model with torsion 
\begin{equation}
\Omega_m+\Omega_k+\Omega_{\alpha}=1,
\end{equation}
with the matter density parameter $\Omega_m=\frac{\kappa\rho_{m}}{3H^2}$, the curvature density parameter  $\Omega_k\equiv -\frac{k}{a^2 H^2}$, and the torsion density parameter $\Omega_{\alpha}=4\alpha-4 \alpha^2$. For convenience, we denote the $\Omega_{m}$ and $\Omega_{k}$ as the present density parameters in the subsequent analysis, respectively.} Now Eq.~(\ref{H2case2}) could be rewritten as 
\begin{equation}
    H^2(z;\mathbf{p})= H_0^2\bigg(\frac{\Omega_{m}(1+z)^{3-2\alpha}+\Omega_k(1+z)^2}{1-4\alpha+4\alpha^2}\bigg)\, , \label{11}
\end{equation}
where $\Omega_k=1-\Omega_{m} -4\alpha+4\alpha^2$ and the redshift is introduced by $(1+z)=a_0/a$. The parameter $\mathbf{p}$ denotes relevant cosmological model parameters, i.e., $\mathbf{p}\,=\,[H_0,\Omega_m, \alpha]$ for flat cosmological model and $\mathbf{p}\,=\,[H_0,\Omega_m, \Omega_k, \alpha]$ for non-flat cosmological model. Due to the normalization condition, there are only two degrees of freedom in the model in the flat case and three degrees of freedom in the non-flat case.
\subsection{Case II: $\phi(t)=-H(t)/2$  with non-vanishing cosmological
constant}

For $\phi(t)=-\alpha H(t)$ with $\Lambda\neq0$ and $\omega=0$, one could obtain the Friedmann equation as
 \begin{equation}
        H^2 =\kappa\rho_m  - \frac{k}{a^2}+ \frac{\Lambda}{3} +4\alpha H^2-4\alpha^2 H^2\,,\label{H2a}
\end{equation}
and the solution for the energy density from is
\begin{eqnarray}
    \rho_m(a) = \Bigg[\rho_{m0} + \frac{4\alpha \Lambda}{3\kappa (3-2\alpha)}\Bigg]\Bigg(\frac{a_0}{a} \Bigg)^{3-2\alpha} - \frac{4\alpha \Lambda}{3\kappa (3-2\alpha)}\,.\label{sol22}
\end{eqnarray}
For $\alpha = 0$ and $\Lambda \neq0$ the evolution of the energy density is exactly the expected one, namely $ \sim a^{-3}$.

\textbf{Similar to Case I, we introduce the density parameter for cosmological constant $\Omega_{\Lambda}=\frac{\Lambda}{3H^2}$. For convenience, we also denote $\Omega_{\Lambda}$ as the present value of cosmological constant density parameter. Based on the  conservation equation of 
\begin{equation}
\Omega_m+\Omega_k+\Omega_{\Lambda}+\Omega_{\alpha}=1,
\end{equation}
}
the Friedmann equation can be written as
\begin{equation}
    H^2(z;\mathbf{p}) = {H_0^2}\frac{1}{(1-2\alpha)^2}\Bigg[\Big(\Omega_{m}+\frac{4\alpha}{(3-2\alpha)}\Omega_\Lambda\Big)(1+z)^{3-2\alpha} + \Omega_k(1+z)^2+\Omega_{\Lambda}\Big(1-\frac{4\alpha}{3-2\alpha}\Big)\Bigg]\,,\label{14}
\end{equation}
where $\mathbf{p}\,=\,[H_0,\Omega_m, \alpha]$ for flat cosmological model and $\mathbf{p}\,=\,[H_0,\Omega_m, \Omega_k, \alpha]$ for non-flat cosmological model. By using the Friedmann eqaution at $z=0$, namely $1=(\Omega_m + \Omega_k + \Omega_{\Lambda})/(1-2\alpha)^2$, we can express $\Omega_\Lambda $ as a function of $\Omega_{m}$, namely $\Omega_\Lambda = (1-2\alpha)^2 - \Omega_m-\Omega_k$. Thus, there are three degrees of freedom for the model in the flat case and four degrees of freedom in the non-flat case. From this equation, we can see that the standard cosmological constant term $\Omega_\Lambda$ is also affected by the torsion coupling, which nest to the standard model when $\alpha \to 0$.

\section{High-redshift observations of quasars and BAO}

As the brightest sources in the Universe, quasars have considerable potential to be used as useful cosmological probes \citep{Risaliti2015,Risaliti2017,LiuT20a,LiuT20b}. Although the extreme variability in their luminosity and high observed dispersion, many efforts have been made to standardize quasars as probes based on their own properties, such as the Baldwin effect \citep{Baldwin77}, the Broad Line Region radius - luminosity relation \citep{Watson11}. Recently, \citet{Risaliti2018} attempted to use quasars as standard candles by using the nonlinear relation between their intrinsic UV and the X-ray luminosities. So far, the largest quasar sample with both X-ray and UV observations consists of $\sim 12,000$ objects. Through the gradually refined selection technique and flux measurements, as well as the elimination of systematic errors caused by various aspects, 2421 quasars covering the redshift range $0.009<z<7.5$ were left in the final cleaned sample \citep{Lusso:2020pdb}. The scatter plot of the final quasar 
sample is shown in the left panel of Fig.~1 and we denote it as "QSO[UVX]" in the following analysis.

The non-linear relation between luminosities in the X-rays ($L_X$) and UV band ($L_{UV}$) is given by
\begin{equation} \label{X-UV relation}
\log_{10}(L_X)=\gamma\log_{10}(L_{UV})+\beta',
\end{equation}
where $L_X$ and $L_{UV}$ represent the monochromatic luminosities at 2keV and 2500${\AA}$ with rest-frame, while $\beta'$ and $\gamma$ denote the intercept and the slope parameters. Such relation can be rewritten in terms of the fluxes $F$, the slope $\gamma$ and the normalization constant $\beta$ as
\begin{equation} \label{16}
\log_{10}(D_L(z))=\frac{1}{2(1-\gamma )}\times[\gamma\log_{10}(F_{UV})-\log_{10}(F_X)
+ \beta ],
\end{equation}
where $D_L(z)$ is luminosity distance to the quasar, $\beta$ is a constant that quantifies the slope and the intercept, $\beta=\beta'+(\gamma-1)\log 4\pi$. We also quantify the intrinsic dispersion of such relation with a parameter $\delta$, which is not shown directly in the equation. However, the crucial question is what is the slope and intercept parameters of the quasar source? 
More recently, \citet{2021MNRAS.507..919L} estimated that $\beta = 7.735 \pm 0.244$, $\gamma = 0.649 \pm 0.007$, and $\delta = 0.235 \pm 0.04$ by model-independently reconstructing the expansion history of the Universe from the latest observations of type Ia supernova. In this analysis, we adopt the calibration results of these quantities.

For the second type of quasar data, we focus on the ultra-compact structure in radio quasars with milliarcsecond angular sizes measured by very-long-baseline interferometry (VLBI) \cite{Kellermann93}. The angular size of a compact radio quasar is defined as $\theta={2\sqrt{-\ln\Gamma \ln 2} \over \pi B}$,
where $B$ is the interferometer baseline measured in wavelengths and the visibility modulus $\Gamma=S_c/S_t$ is the ratio between the total and correlated flux densities. Based on a simple relation between the angular size and cosmological distance, the angular diameter distance can be written as
\begin{equation}\label{17}
D_A(z)=\frac{l_m}{\theta(z)},
\end{equation}
where $D_A(z)$ is the angular diameter distance at redshift $z$ , $l_m$ is the intrinsic linear size of the source, and $\theta(z)$ is the observed angular size measured by VLBI. Based on 1398 detected candidates and 917 selected sources from an updated 2.29 GHz VLBI survey (http://nrl.northumbria.ac.uk/13109/), \citet{Cao:2017ivt} recompiled the dataset of 120 intermediate-luminosity radio quasars ($10^{27}W/Hz < L < 10^{28} W/Hz$) with reliable measurements of  angular size of the compact raido quasars. In the previous work, \citet{Cao2017JCAP} demonstrated that the intrinsic linear size $l_m$ shows negligible dependence on both redshifts and intrinsic luminosity. Thus, these ultra-compact radio quasars are available as standard rulers for cosmology, which is extended to multi-frequency VLBI observations in the subsequent studies \cite{Cao:2017abj}. The detailed angular sizes $\theta(z)$ and redshifts $z$ data listed in Table 1 of \citet{Cao:2017ivt}, which extend the Hubble diagram to a redshift range $0.46<z<2.76$.  We denote this sample as "QSO[RS]" in this work. We remark hert that the intrinsic linear size $l_m$ also needs to be calibrated with external indicators such as SNe Ia. In this analysis, we adopt the calibration results of such quantity $l_m=11.28\pm0.25$ pc through a new cosmology-independent technique \citep{Cao2017JCAP}. This results and dataset have been extensively used for cosmological applications in the literature \citep{Qi17,Ma17,Cao19,Liuplb,Liuepjc,LiuMN}. In the right panel of Fig.~1, we show the scatter plot of the 120 intermediate-luminosity radio quasars sample.

Finally, the matter clustering of Baryon Acoustic Oscillation (BAO) can be considered as a standard ruler for length scale in cosmology. These  signals can be used to measure the Hubble parameter $H(z)$ and angular diameter distance in the radial and tangential directions, respectively. The BAO data are derived from large-scale structural power spectra of astronomical observations. In the previous work \cite{2021MNRAS.505.2111L}, the authors indicated that the results show that the combined QSO data alone are not able to provide tight constraints on model parameters, which is mainly related to the large dispersion. In addition, they demonstrated that the BAO data is complementary to the QSO distance measurements. Thus, in order to get more stringent constraints on cosmological parameters, we also consider 11 recent BAO measurements from BOSS DR12 at $z=0.38, 0.51, 0.61$, 6dFGs and SDSS MGS at $z=0.122$, DES Y1 results at $z=0.81$, and eBOSS DR14 at $z=1.52, 2.34$, which are summarized in Table 1 of \citet{2021MNRAS.505.2111L}.

\begin{figure*}
\begin{center}
{\includegraphics[width=0.48\linewidth]{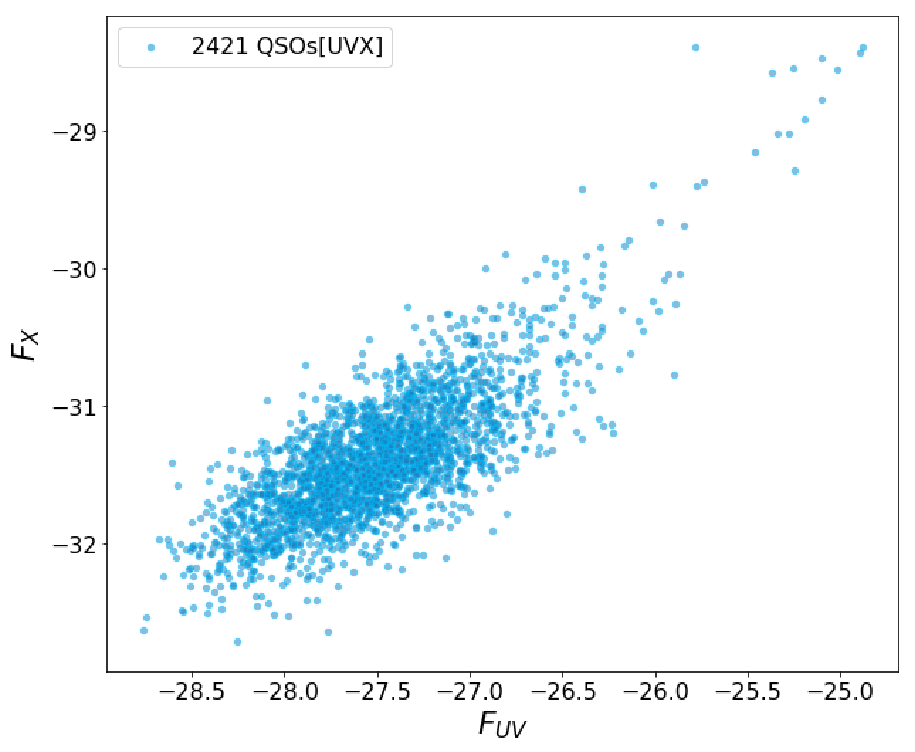}
\includegraphics[width=0.48\linewidth]{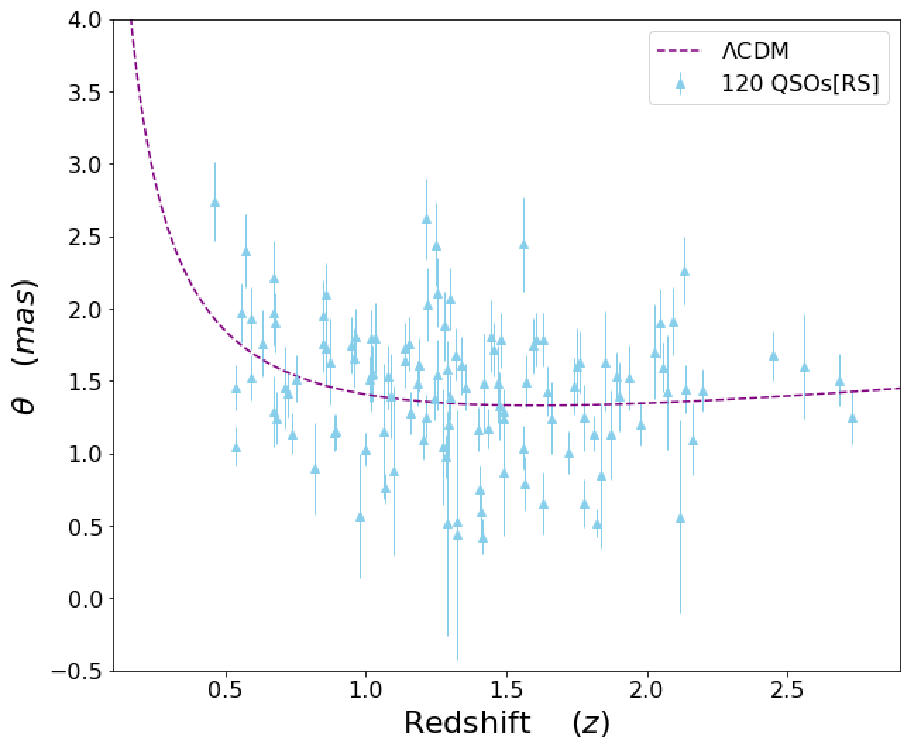}}
\end{center}
\caption{\textit{Left panel:} The scatter plot of UV and X-ray fluxes of 2421 quasars. \textit{Right panel:} The scatter plot of angular sizes of 120 compact radio quasars. The purple dotted line denotes the angular sizes calculated from the fiducial flat $\Lambda$CDM model ($H_0=70.0$ km/s/Mpc, $\Omega_m=0.30$). }
\end{figure*}

\section{Methodology and results}

In the framework of FRLW metric, the luminosity distance $D_L(z)$ and angular diameter distance $D_A(z)$ can be expressed as
\begin{equation}\label{18}
D_L(z)=\frac{D_A(z)}{(1+z)^2} = \left\lbrace \begin{array}{lll}
\frac{(1+z)}{\sqrt{|\Omega_{\rm k}|}}\sinh\left[\sqrt{|\Omega_{\rm k}|}\int_{0}^{z}\frac{dz'}{H(z')}\right]~~{\rm for}~~\Omega_{k}>0,\\
(1+z)\int_{0}^{z}\frac{dz'}{H(z')}~~~~~~~~~~~~~~~~~~{\rm for}~~\Omega_{k}=0, \\
\frac{(1+z)}{\sqrt{|\Omega_{\rm k}|}}\sin\left[\sqrt{|\Omega_{\rm k}|}\int_{0}^{z}\frac{dz'}{H(z')}\right]~~~~{\rm for}~~\Omega_{k}<0.\\
\end{array} \right.
\end{equation}
Here $H(z)$ is the cosmic expansion rate at redshift $z$, which 
given by Eq. (\ref{11}) for Case I and Eq. (\ref{14}) for Case II in torsion cosmology.

In order to determine the cosmological parameters $\mathbf{p}$ in different models, we use Monte Carlo Markov Chain (MCMC) method to achieve our purpose. The chi-square $\chi^2$ function for the QSO[UVX] sample is defined as 
\begin{eqnarray}
\mathcal{L}_{QSO[UVX]}=\prod_{i=1}^{2421}\frac{1}{\sqrt{2\pi\sigma_{i}^2}}
\times\exp\bigg\{
-\frac{\left[D_L^{th}(z_i;\mathbf{p})
     - D^{obs}_{L, i}\right]^{2}}{2\sigma_{i}^{2}}\bigg\},
\end{eqnarray}
where $D^{obs}_{L, i}$ is the luminosity distance of $i-$th QSO[UVX] data (see Eq.~(\ref{16})), with $\sigma_{i}$ representing its uncertainty, and the $D_L^{th}$ is theoretical counterpart expressed in terms of cosmological parameters.
For the QSO[RS] sample, the chi-square $\chi^2$ function is given by
\begin{eqnarray}
\mathcal{L}_{QSO[RS]}=\prod_{i=1}^{120}\frac{1}{\sqrt{2\pi\sigma_{i}^2}}
\times\exp\bigg\{
-\frac{\left[D_A^{th}(z_i;\mathbf{p})
     - D^{obs}_{A, i}\right]^{2}}{2\sigma_{i}^{2}}\bigg\},
\end{eqnarray}
where $D^{obs}_{A, i}$ is the angular-diameter distance of $i-$th QSO[RS] data (see Eq. (\ref{17})), with $\sigma_{i}$ representing its uncertainty, and the $D_A^{th}$ is theoretical counterpart expressed in terms of cosmological parameters. The distance uncertainties are obtained by the standard propagation equation from the uncertainties of different observables.
 
For the use of BAO data, we follows the approach carried out in \citet{2021MNRAS.505.2111L}, considering the possible correlation between different BAO observations. Specially, for the three uncorrelated BAO measurements \cite{BAO0,BAO01,BAO02}, the corresponding log-likelihood function is given by
\begin{equation}
\mathrm{ln}({\cal L}_{BAO})=-\frac{1}{2}\sum^3\frac{[A_{th}(z_i;\mathbf{p})-A_{obs}]^2}{\sigma^2}.
\end{equation}
For the correlated BAO measurements, these covariances are publicly available in refs \cite{BAO1,BAO2,BAO3}. The log-likelihood function of correlated BAO measurements takes the form
\begin{equation}
\mathrm{ln}({\cal L}_{BAO})={[A_{th}(\mathbf{p})-A_{obs}]^T}C^{-1}{[A_{th}(\mathbf{p})-A_{obs}]},
\end{equation}
where $C^{-1}$ is the the inverse covariance matrix, while $A_{th}$ and $A_{obs}$ are the predicted and measured quantities of the BAO data listed in the Table 1 of \citet{2021MNRAS.505.2111L}.

The total log-likelihood
\begin{equation}
\mathrm{ln} {\cal L}= \mathrm{ln} ({\cal L}_{\rm{QSO[RS]}})+\mathrm{ln}
({\cal L}_{QSO[UVX]})+\mathrm{ln}
({\cal L}_{BAO})
\end{equation}
is sampled in the framework of Python MCMC module \textit{emcee} \citep{Foreman_Mackey_2013}. We employ a uniform prior on $H_0$ in the range [0, 150] $\Mpc$, $\Omega_m$ in the range [0.05, 0.5],  $\Omega_k$ in the range [-1, 1] (if available), and torsion parameter $\alpha$ in the range [-0.5, 0.5] to make sure that they are either physically motivated or sufficiently conservative.

\begin{figure*}
\begin{center}
{\includegraphics[width=0.45\linewidth]{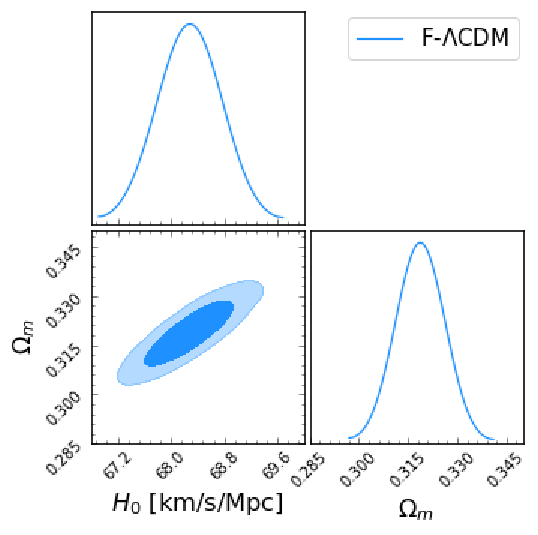}
\includegraphics[width=0.5\linewidth]{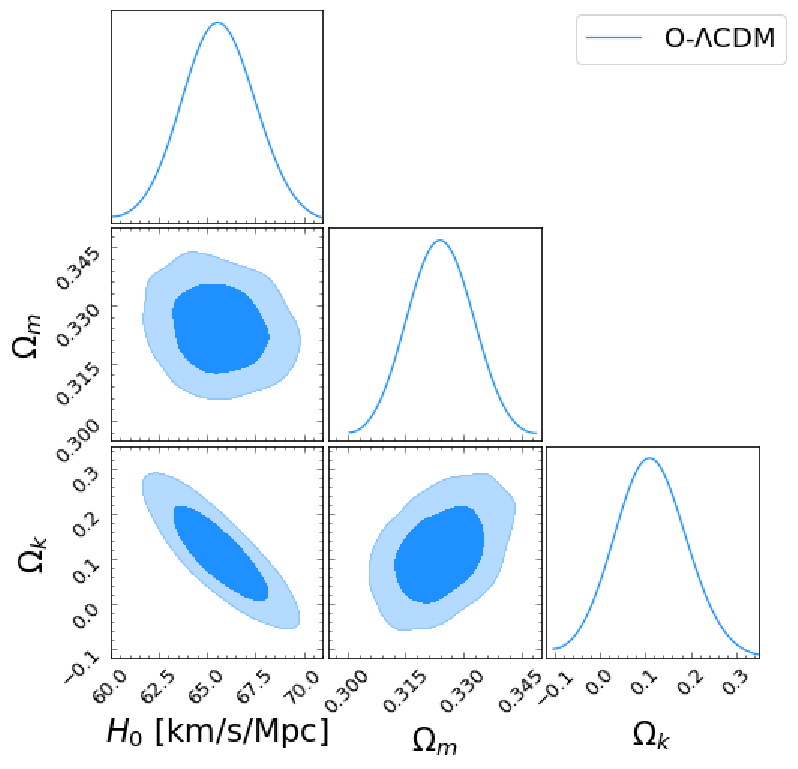}}
\end{center}
\caption{Two-dimensional and marginalized distributions of the 
 cosmological parameters $H_0$, $\Omega_m$ and $\Omega_k$ (if available) in flat $\Lambda$CDM (left panel), and non-flat $\Lambda$CDM models (right panel) with QSO[UVX], QSO[RS], and BAO datasets.}
\end{figure*}

In order to demonstrate the constraining power of such observational datasets, we will discuss $\Lambda$CDM, Case I, and Case II models both in flat and non-flat cases. For $\Lambda$CDM models, the posterior one-dimensional (1-D) probability distributions and two-dimensional (2-D) confidence regions of the cosmological parameters for flat and non-flat $\Lambda$CDM models are shown in Fig. 2.  We obtain the best-fit values with 68.3\% confidence level for the cosmological parameters: $H_0=68.28^{+0.40}_{-0.41}$ $\Mpc$ and $\Omega_m=0.319^{+0.006}_{-0.006}$ in the framework of flat  $\Lambda$CDM model. Our constraints on $H_0$ and $\Omega_m$ in the framework of flat $\Lambda$CDM cosmology are well consistent with the recent results of Planck collaboration within 1$\sigma$ confidence level. Such combination of datasets results in an precision on Hubble constant $\Delta H_0 \sim 0.5$  and matter density $\Delta \Omega_m\sim10^{-3}$ in the flat standard cosmological model. In non-flat $\Lambda$CDM model, the best-fit values are  $H_0=65.59^{+1.53}_{-1.46}$ $\Mpc$, $\Omega_m=0.324^{+0.007}_{-0.007}$, and $\Omega_k=0.110^{+0.065}_{-0.062}$. Although our constraint results slightly favor a open universe, there is no obvious signal indicating the deviation from a flat Universe within $2\sigma$ confidence level. Such unambiguous result by combining quasars and BAO datasets is also consistent with the current analysis results which inferred $\Omega_k$ through other model-independent methods \cite{LiuT20b,2020ApJ...901..129L,2021MNRAS.503.2179Q}. The numerical results are summarized in Table 1.

\begin{figure*}
\begin{center}
{\includegraphics[width=0.45\linewidth]{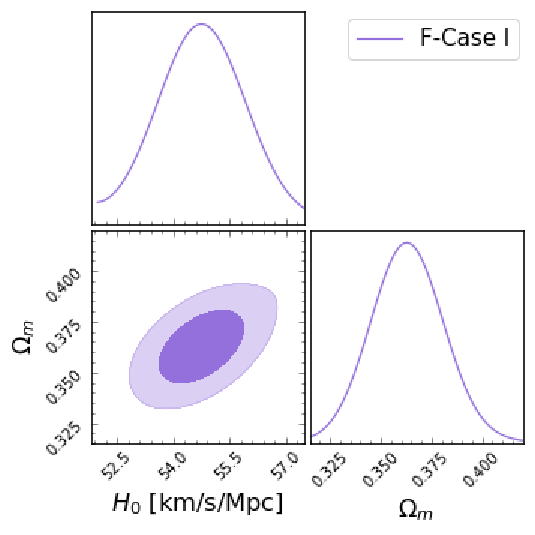}
\includegraphics[width=0.5\linewidth]{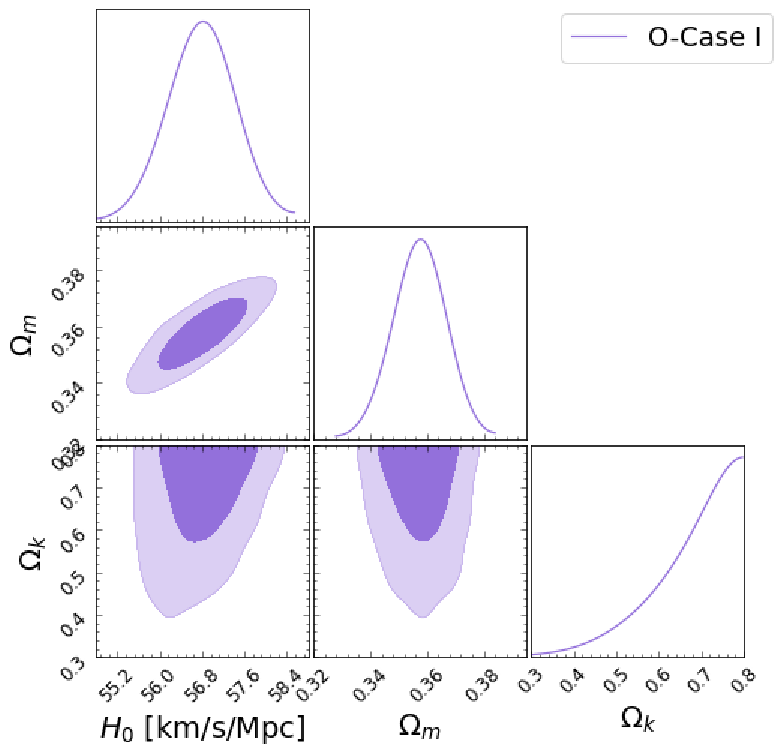}}
\end{center}
\caption{Two-dimensional and marginalized distributions of the 
 cosmological parameters $H_0$, $\Omega_m$ and $\Omega_k$ (if available) for Case I cosmology with torsion.}
\end{figure*}

Focusing on Case I, i.e., the torsion term leads to the accelerated expansion of the Universe without cosmological constant $\Lambda$. For the flat Case I, we obtain a lower Hubble constant $H_0=54.75^{+0.55}_{-0.48}$ $\Mpc$ and a higher matter density $\Omega_m=0.363^{+0.009}_{-0.008}$ from high-redshift observational data. The left panel of the Fig. 3 shows the 1$\sigma$ and 2$\sigma$ contours for the parameters $H_0$ and $\Omega_m$. The best-fit value of $H_0$ is in disagreement with the result inferred from Planck 2018 observation and local measurement from SH0ES collaboration. This result is likely due to strong degeneracy between the Hubble constant and matter density. It is worth noting that the torsion parameter $\alpha$ has no extra degrees of freedom, and relates with the matter density parameter $\Omega_m$, due to the normalization condition $\Omega_{m}+4\alpha-4\alpha^2=1$ in this case. Thus, we 
obtain $\alpha=0.198^{+0.004}_{-0.004}$, which is inconsistent with the previous results indicating a small value of $\alpha$ supported by the low-redshift Hubble parameter and SN Ia observations \cite{torsion1,torsion2,torsion3}. In the framework of non-flat Case I model, one can clearly see that the right panel of the Fig. 3 shows that the combination of quasars and BAO datasets provides stringent constraints on $H_0=56.81^{+0.490}_{-0.534}$ $\Mpc$ and $\Omega_m=0.358^{+0.008}_{-0.008}$. More interestingly, the best-fit value of $\alpha=-0.017^{+0.033}_{-0.018}$ is perfectly consisted with the result by studying the torsion effect on the primordial helium-4 abundance \cite{Kranas2019}, where a narrow interval for the $\alpha=-\lambda$ parameter was found $(-0.0058 <\lambda< +0.0194)$. The numerical results can also be found in Table 1. 

\begin{table*}\renewcommand\arraystretch{1.4}
\centering
\begin{tabular}{lcccccc}
\hline
\hline
 &F-$\Lambda$CDM & O-$\Lambda$CDM& F-Case I  & O-Case I   & F-Case II   & O-Case II  \\
\hline
{\boldmath$H_0$}& $68.28^{+0.40}_{-0.41}$ & $65.59^{+1.53}_{-1.46}$  & $54.75^{+0.55}_{-0.48}$ & $56.81^{+0.490}_{-0.534}$ & $66.09^{+1.30}_{-1.36}$ & $66.16^{+1.79}_{-1.61}$\\
\hline
{\boldmath$\Omega_m$} & $0.319^{+0.006}_{-0.006}$ & $0.324^{+0.007}_{-0.007}$ & $0.363^{+0.009}_{-0.008}$ & $0.358^{+0.008}_{-0.008}$ & $0.321^{+0.007}_{-0.007}$ & $0.322^{+0.010}_{-0.011}$ \\
\hline
{\boldmath$\Omega_k$}& $-$ & $0.110^{+0.065}_{-0.062}$  & $-$ & $0.716^{+0.066}_{-0.123}$ & $-$ &$0.030^{+0.285}_{-0.217}$ \\
\hline
{\boldmath$\Omega_{\alpha}$}& $-$ & $-$ & {\boldmath$0.637^{+0.010}_{-0.010}$} & {\boldmath$-0.074^{+0.132}_{-0.067}$}& {\boldmath$0.097^{+0.063}_{-0.056}$} & {\boldmath$0.063^{+0.200}_{-0.245}$} \\
\hline
{\boldmath$\Omega_{\Lambda}$}& {\boldmath$0.681^{+0.006}_{-0.006}$} & {\boldmath$0.566^{+0.069}_{-0.072}$} & $-$& $-$& {\boldmath$0.582^{+0.063}_{-0.070}$} & {\boldmath$0.585^{+0.473}_{-0.495}$} \\
\hline
{\boldmath$\alpha$} & $-$ & $-$ & $0.198^{+0.004}_{-0.004}$ & $-0.017^{+0.033}_{-0.018}$ & $0.025^{+0.017}_{-0.015}$ & $0.016^{+0.055}_{-0.064}$\\
\hline
{\boldmath$\chi^2_{min}$}& $2999.04$ & $2996.15$  & 3067.48 & 3037.81 & 2996.35 & 2996.38\\
\hline
{\boldmath$\chi^2_\nu$}& 1.175& $ 1.174$  & 1.202 & 1.190 & 1.174 &  1.174\\
\hline
{\boldmath$p$} & 2 & 3 & 2 & 3 & 3 & 4 \\
\hline
AIC & $3003.04$ & $3002.15$ &3071.48 & 3041.81 & 3002.35 & 3004.38 \\
\hline
BIC & $3005.86$ & $3006.36$ & 3074.30 & 3048.04 &  3006.58 & 3010.01\\
\hline
\hline
\end{tabular}
\caption{Fits on flat $\Lambda$CDM (denoted as F-$\Lambda$CDM), non-flat $\Lambda$CDM (denoted as O-$\Lambda$CDM) models, flat Case I (denoted as F-Cases I), non-flat Case I (denoted as O-Cases I), flat Case II (denoted as F-Cases II), non-flat Case II (denoted as O-Cases II) models from combined quasar and BAO data. and LD data. We also give the values of minimum $\chi_{min}^2$, relative chi-square $\chi_{\nu}^2$, Akaike Information Criterion (AIC), and Bayesian Information Criterion (BIC).}
\end{table*}

In the framework of flat and non-flat Case II models, i.e., torsion term and cosmological constant $\Lambda$ term simultaneously drive the accelerated expansion of the Universe. The left panel of Fig. 4 shows the constraint results ($1\sigma$ and $2\sigma$ contours) for $H_0$, $\Omega_m$, and $\alpha$ for the flat Case II model, by using 
the observations of high-redshift quasars and BAO. Table I presents the best-fit values of the parameters with $1\sigma$ uncertainties. We see that both $H_0=66.09^{+1.30}_{-1.36}$ $\Mpc$ and $\Omega_m=0.321^{+0.007}_{-0.007}$ are well consistent with the results of Planck 2018 CMB observations based on $\Lambda$CDM model, with a small positive value of the torsion term of $\alpha=0.025^{+0.017}_{-0.015}$. Such model based on these parameters is fully compatible with the scenario that cosmic acceleration being represented by the torsion parameter $\alpha$ and the cosmological constant $\Lambda$. For the non-flat Case II model, the final results of cosmological parameters are $H_0=66.16^{+1.79}_{-1.61}$ $\Mpc$, $\Omega_m=0.322^{+0.010}_{-0.011}$,  $\Omega_k=0.030^{+0.285}_{-0.217}$, and $\alpha=0.016^{+0.055}_{-0.064}$. The numerical results are summarized in Table 1 and presented in the right panel of Fig. 4. Our constraints on cosmological parameters are well consistent with the recent study of \citet{2021MNRAS.505.2111L}, in which several modified gravity theories are probed with multiple measurements of high-redshift quasars and BAO datasets. It is worth noting that the matter density parameter $\Omega_m$ derived from such dataset combination tends to be higher than that from other cosmological probes, as was remarked in the previous works \cite{Risaliti2018,Khadka}. This suggests that the composition of the Universe characterized by cosmological parameters can be comprehended differently through high-redshift quasars. However, 
the determination of $\Omega_k$ suggests no significant deviation from flat spatial hypersurfaces, although favoring a somewhat positive value in the non-flat Case II model. In this case, our results also show that there is no obvious evidence against the existence of the torsion effect in the Universe.

\begin{figure*}
\begin{center}
{\includegraphics[width=0.48\linewidth]{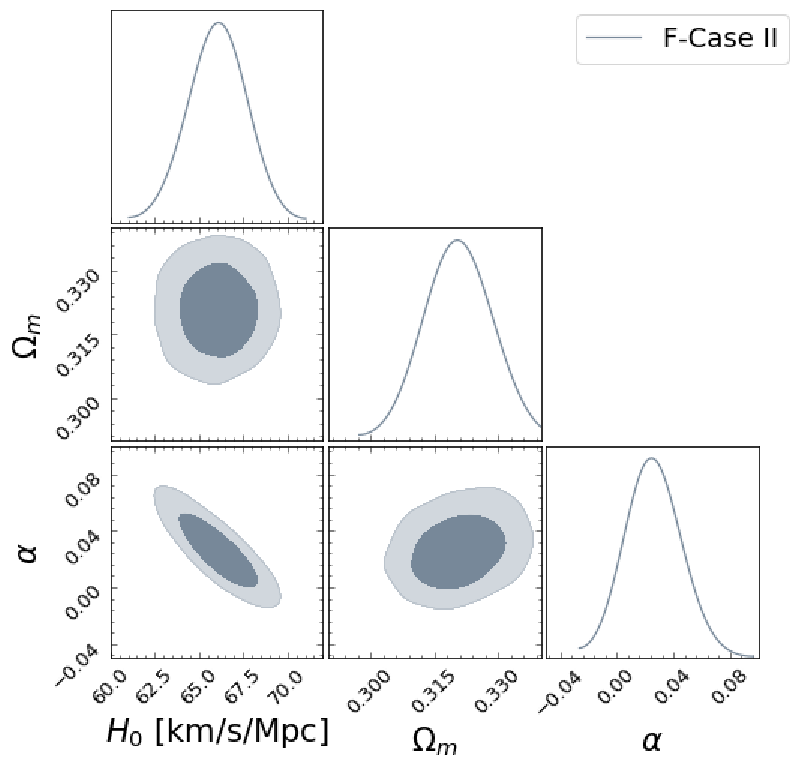}
\includegraphics[width=0.48\linewidth]{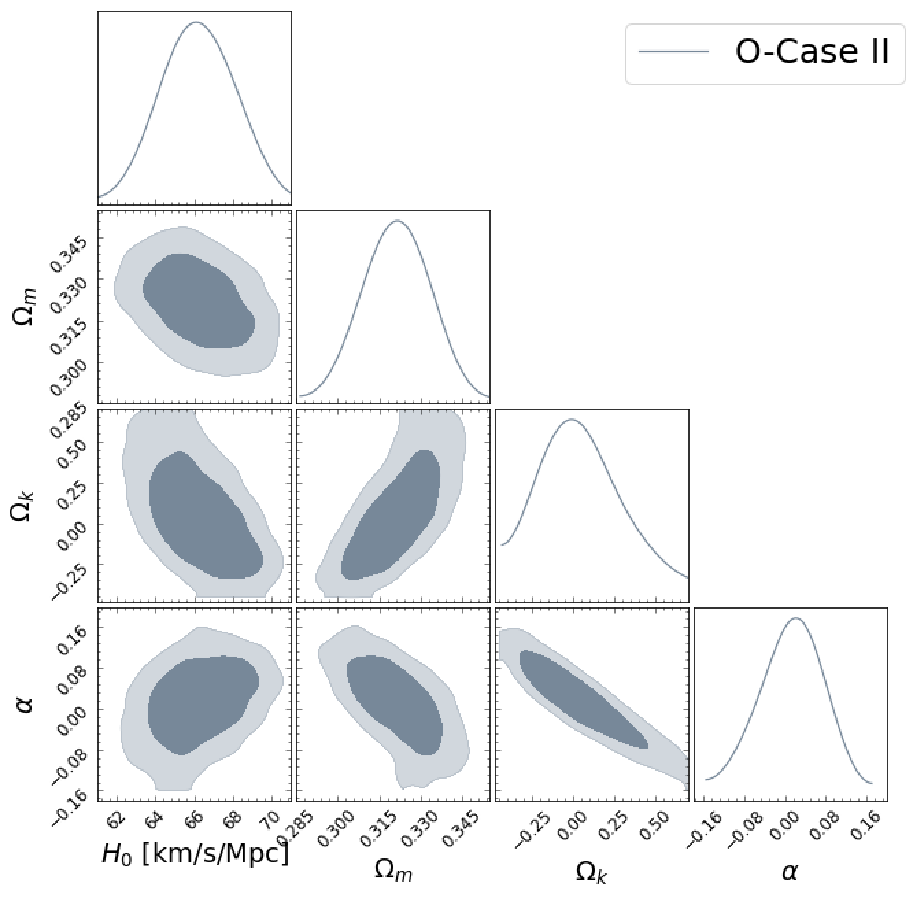}}
\end{center}
\caption{Two-dimensional and marginalized distributions of the 
 cosmological parameters $H_0$, $\Omega_m$ and $\Omega_k$ (if available) for Case II cosmology with torsion. }
\end{figure*}

\textbf{In view of the motivation of this work, it is necessary to 
make some remarks as follows. For the flat and non-flat Case I models, we obtain the best-fit values of the torsion density parameters $\Omega_{\alpha}=0.637^{+0.010}_{-0.010}$ and $\Omega_{\alpha}-0.074^{+0.132}_{-0.067}$, respectively. For the flat and non-flat Case II models, the torsion density parameters are $\Omega_{\alpha}=0.097^{+0.063}_{-0.056}$ and $\Omega_{\alpha}=0.063^{+0.200}_{-0.245}$. Therefore, the accelerating expansion of late-time universe could be well explained by the torsion effect, based on the latest observations of high-redshift data.} To evaluate the complexity  of statistical models and measure the goodness of fit data, we also apply the well-known Information Criteria (IC) to all of the models considered. The corresponding minimum $\chi_{min}^2$, relative chi-square ($\chi^2_\nu=\chi^2_{min}/(n-p)$), Akaike Information Criterion ($\mathrm{AIC}=\chi^2_{min}+2p$), and Bayesian Information Criterion ($\mathrm{BIC}=\chi^2_{min}+p\ln N$) are reported in Table 1, where $n$ is the number of data and $p$ is the number of free parameters. For flat cosmological models, we find that the flat Case II model has the smallest AIC, while the flat $\Lambda$CDM model shows the smallest BIC value BIC. We also note that the differences of AIC and BIC between the flat Case II and flat $\Lambda$CDM model are less than 2, which means these two models are comparable to each other according to both information criteria. For non-flat cosmological models, the non-flat $\Lambda$CDM model is the most preferred model according to both AIC and BIC criteria. The non-flat Case II model, which is a little bit worse than non-flat $\Lambda$CDM, also performs well in explaining the QSO+BAO data. The most disfavored model is flat Case I model, with strong evidence against such cosmological scenario among these models.  We also plot the graphical representation of the AIC and BIC results in Fig.~5, which directly shows the results in the IC test for each model. We conclude that, out of all the candidate models, the $\Lambda$CDM and Case II models are the two most favored cosmological models, which demonstrates that the observations of high-redshift universe may support the existence of the torsion effect. For all of the models considered in this analysis, our results show that there is still some possibility that the standard cold dark matter ($\Lambda$CDM) scenario might not be the best cosmological model preferred by the current observations.

\begin{figure}
\begin{center}
{\includegraphics[width=0.7\linewidth]{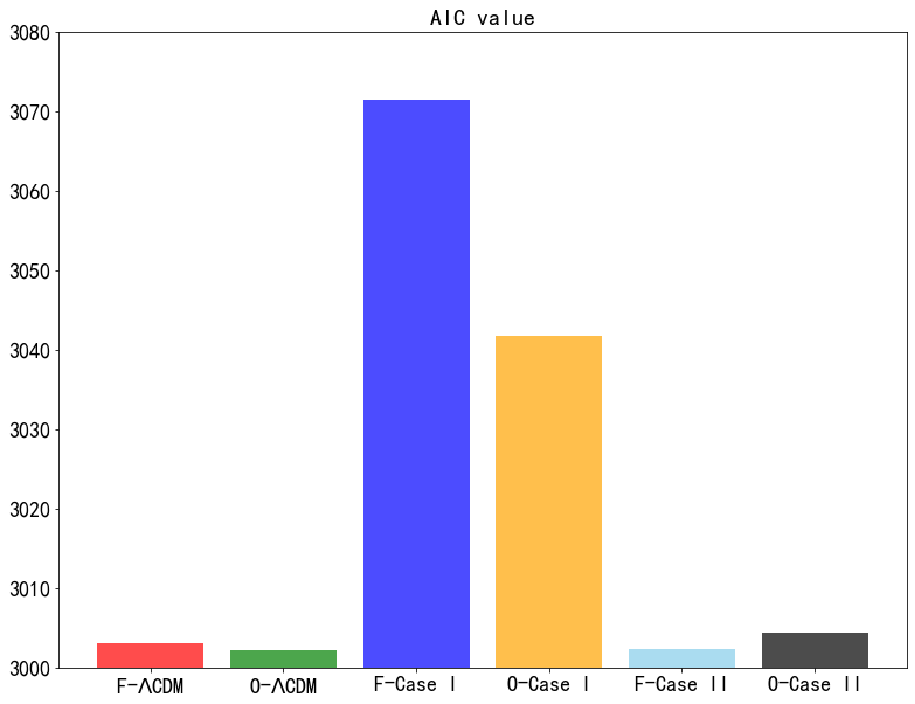} \\
\includegraphics[width=0.7\linewidth]{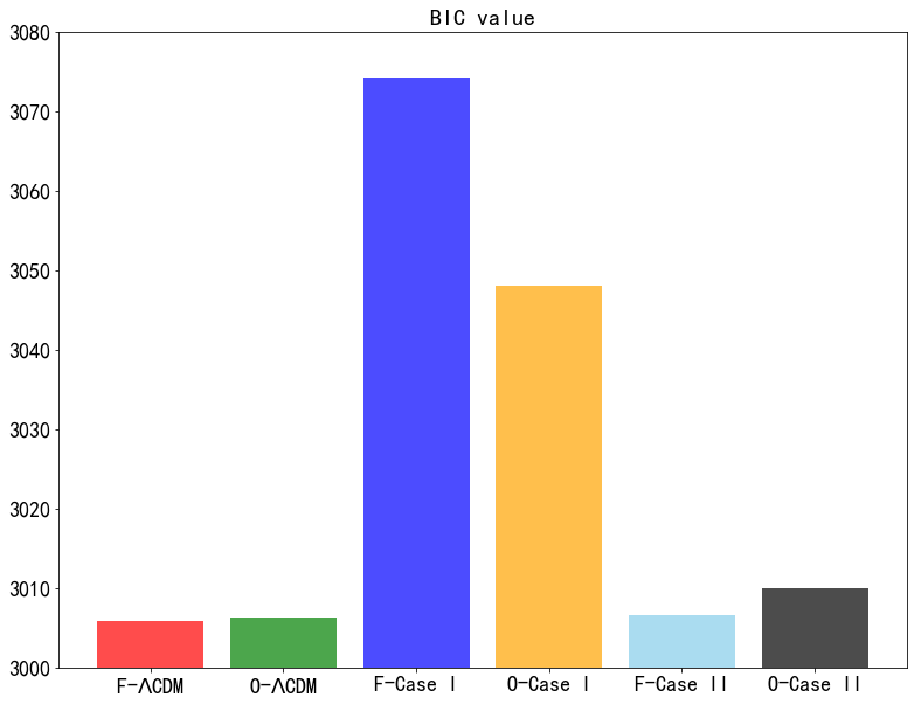}}
\end{center}
\caption{Graphical representation of the results of AIC and BIC for for each model. The order of models from left to right is the same as that in Table I. }
\end{figure}

\section{Conclusion and discussion}

In this paper, we study the Friedmann-like cosmology with torsion effect by using high-redshift observations, where torsion is described by the antisymmetric part of the non-Riemannian affine connection that can drive the accelearated expansion of the Universe. The torsion effect is regarded as a macroscopic manifestation of the intrinsic angular momentum (spin) of matter. Two special cases are investigated, i.e., torsion effect with vanishing cosmological constant and torsion effect with cosmological constant. This study is strongly motivated by the need of revisiting the Hubble constant, spatial curvature and other model parameters in the framework of different cosmological models of interest, and searching for implications for the non-flat Universe and extensions of the standard cosmological model (the spatially flat $\Lambda$CDM).

The latest observations of high-redshift data come from multiple measurements of quasars and baryon acoustic oscillations. Such newly compiled quasar datasets in the cosmological analysis is crucial to this aim, since it will extend the Hubble diagram to high-redshift range in which predictions from different cosmologies can be distinguished. The newest large quasar samples acting as standard distance indicators provide a good opportunity to test models in the "redshift desert" ($0.009<z<7.5$) that is not yet widely available through other observations. In order to improve the cosmological constraints, the quasar datasets are also combined with recent observations of BAO data. For all of the cosmological models considered, our results suggest that there is still some possibility that the standard cold dark matter scenario may not be the best cosmological model preferred by the high-redshift observations. We conclude that out of all the candidate models, the $\Lambda$CDM model and the torsion plus cosmological constant model are strongly 
supported by the current high-redshift data. This indicates that the observations of high-redshift universe may support the existence of the torsion effect, which would be expected to yield the late-time cosmic acceleration. Specially, in the framework of Friedmann-like cosmology with torsion, the determined Hubble constant is in very good agreement with that derived from the Planck 2018 CMB results. On the other hand, our results are compatible with zero spatial curvature and there is no significant deviation from flat spatial hypersurfaces. Finally, we check the robustness of high-redshift observations by placing constraints on the torsion parameter $\alpha$, which is strongly consistent with other recent works focusing on torsion effect on the primordial helium-4 abundance.

As a final remark, there are many potential ways to improve our results. From the observational point of view, current COSMOS survey \cite{obs1}, XMM-Newton Serendipitous Source Catalog Data Release \cite{obs2} and future the Sloan Digital Sky Survey (SDSS) and XMM spectra \cite{obs3} will generate large numbers of quasars that are better controlled for systematics and with significantly lower dispersion. Therefore, it is reasonable to expect that our approach will play an increasingly important role in high-precision studies of
the torsion effect. From the theoretical point of view, the search of other more reliable and reasonable alternative models are also one of the important ways to solve the current predicament of modern cosmology. Some theories suggest that dark matter self-interactions could provide the accelerated expansion of the Universe without any
dark energy component \cite{Cao11AA}. Thinking of dark energy and dark matter as a fluid called Chaplygin gas could provide another possible solution to unify two uncharted territories \cite{Zheng22EPJC}. Distinguishing between physical and geometric models of dark energy, as well as modifying theories of gravity and breaking degeneracy, is actually a fundamental task of modern cosmology. Our work could help us better understand these important issues.

\textbf{Acknowledgements:}

Liu. T.-H was supported by National Natural Science Foundation of China under Grant No. 12203009; Chutian Students of Hubei Chutian Scholars Program; Cao. S was supported by the National Natural Science Foundation of China under Grants Nos. 11690023, 11920101003 and A2020205002; the Strategic Priority Research Program of the Chinese Academy of Sciences, Grant No. XDB23000000; the Interdiscipline Research Funds of Beijing Normal University; the China Manned Space Project (Nos. CMS-CSST-2021-B01 and CMS-CSST-2021-A01); and the CAS Project for Young Scientists in Basic Research under Grant No. YSBR-006.

\textbf{Data Availability:} This manuscript has associated data in a data repository. [Authors' comment:The data underlying this paper will be shared on reasonable request to the corresponding author.]

\end{document}